\documentclass[reprint,amsmath,amssymb,aps,pra,showkeys,superscriptaddress,floatfix]{revtex4-2}
\usepackage{gensymb}
\usepackage[utf8]{inputenc}
\usepackage[english]{babel}
\addto\captionsenglish{}
\usepackage{color,epsfig}
\usepackage{array}
\usepackage{amsmath,amssymb}
\usepackage{natbib}
\usepackage{bbold}
\usepackage{dsfont}
\usepackage{graphicx}
\usepackage[normalem]{ulem} 
\usepackage[numbered]{matlab-prettifier}
\usepackage[autostyle]{csquotes}
\usepackage{comment}
\usepackage{float} 
\usepackage{bm}
\usepackage{textcomp}
\usepackage{amsmath}
\usepackage{siunitx}
\begin{document}


\title{A Platform for Evanescently Trapping $^\textrm{87}$Rb Using Silicon Nitride Strip Waveguides Buried in Silica}

\author{S.J. Harding}
\affiliation{Quantum Engineering Technology Labs, H.H. Wills Physics Laboratory and School of Electrical, Electronic, \& Mechanical Engineering, University of Bristol, Tyndall Avenue, BS8 1FD, United Kingdom.}%
\affiliation{Quantum Engineering Centre for Doctoral Training, University of Bristol, Bristol BS8 1FD, United Kingdom}
\author{C.A. Weidner}
\affiliation{Quantum Engineering Technology Labs, H.H. Wills Physics Laboratory and School of Electrical, Electronic, \& Mechanical Engineering, University of Bristol, Tyndall Avenue, BS8 1FD, United Kingdom.}%

\date{\today}

\begin{abstract}
Cold-atom systems have emerged as a highly promising avenue for quantum-enhanced position, navigation, and timing applications. However, their wider adoption is currently hampered in part by the large footprint of the systems. In leveraging the miniaturisation possible through photonic integrated circuits, cold-atom sensors would be able to reach much wider commercial adoption. In this paper, we introduce a platform for evanescently trapping $^\textrm{87}$Rb using strip silicon nitride waveguides buried in silica using red- and blue-detuned fundamental and higher-order modes, providing a three-dimensional adjustable trap for BEC-based, chip-scale work in quantum science and technologies. 

\end{abstract}
\maketitle

\section{Introduction}
Ultracold atoms represent the future of precision measurement of time and acceleration. Since the first magneto-optical trap (MOT) \cite{raab_trapping_1987} and the first experimental observation of a Bose-Einstein condensate (BEC) \cite{anderson_observation_1995}, research on engineering applications of ultracold matter has expanded significantly. Caesium fountain clocks remain the basis for the SI second \cite{bipm_international_2019}, and optical lattice clocks have achieved fractional uncertainties as low as $10^{-18}$ \cite{mcgrew_atomic_2018}. Global navigation satellite systems rely on microwave transitions in rubidium and caesium atoms for timing and positioning and underpin modern navigation, while atom interferometry has enabled high-precision inertial sensing \citep{kasevich_atomic_1991,dickerson_multiaxis_2013,weidner_experimental_2018,rosi_precision_2014} for gravimetry and other emerging applications. 

Just as laser-cooling techniques revolutionised the field, the implementation of atom chips~\cite{keil_fifteen_2016} has allowed for highly localised traps to be formed reproducibly, leveraging mature CMOS-compatible fabrication technology. Permanent \cite{fernholz_dynamically_2007, hannaford_lattices} and current-generated \cite{salim_ultracold_2011} magnetic traps reduce the system's footprint and further reduction to device size, weight, and power are possible with micro-machined vapour cells \cite{dyer_micro-machined_2022} and grating MOTs (gMOTs) \citep{nshii_surface-patterned_2013,cotter_design_2016}, in which a single beam diffracted from a patterned surface provides laser cooling.

More recent work has shown that photonic integrated circuits (PICs) can replace free-space optics for light-matter interactions ~\cite{Hinds}. Through hybrid integration, PIC-based laser sources on Si$_3$N$_4$ (silicon nitride, or SiN) achieve narrow linewidths and stable frequency locking suitable for atom trapping \citep{heim_hybrid_2025,sinclair_14_2020,gallacher_integrated_2019}; support low-loss transmission at near-infrared wavelengths \citep{blumenthal_silicon_2018}; and allow mode conversion, modulation, and splitting \citep{gallacher_silicon_2022}. Alternative waveguides such as fibers \citep{sadgrove_quantum_2016, vetsch_optical_2010} and photonic crystal structures \citep{kim_trapping_2019,douglas_quantum_2015,Kimble,Kimble-2} have proved successful in trapping atoms for interferometric experiments and light-matter interaction. Recent work has also demonstrated the viability of quantum technologies with atoms trapped in the vicinity of integrated photonic structures \cite{xu_dynamics_2025,menon_integrated_2024}. PIC traps using suspended silica have also been suggested \citep{burke_designing_2002,ovchinnikov_perspective_2022,ovchinnikov_optical_2023}, as have comb waveguides~\cite{Fayard2022}. The advantage of SiN based PICs lies in their low-loss, CMOS compatibility~\cite{Mitchell}, and broad transparency window, allowing for atom-photonic integrated technologies to advance dramatically. 

Here, we present a complete PIC-based platform for trapping and manipulation of a $^{87}$Rb BEC aimed at inertial and timing applications. This builds on previous atom chip technology \citep{calviac_bose-einstein-condensate_2025,li_review_2023,chen_planar-integrated_2022,heine_compact_2025}. In particular, we propose to use a gMOT to initially cool the atoms to the Doppler limit before loading the cooled atoms into a chip-scale magnetic trap for evaporative cooling and transport of the atoms to the chip's surface; these are discussed in Sec.~\ref{sec:loading}. Sec.~\ref{sec:PIC} discusses atom trapping using a rectangular strip waveguide of SiN buried in SiO$_2$, showing how one can achieve full three-dimensional confinement through the use of higher order transverse modes and a lattice potential, with the trap's frequency, scattering rate, and lifetime presented. A new challenge with this platform is the requirement to concurrently support modes of different wavelengths and orders in a single waveguide. Potential system failure modes are presented in Sec.~\ref{sec:banana}, where limits on stray charges are established and the effect of chiral fields (which deleteriously affects other atom-trapping platforms) are seen to be not of practical relevance. Together, these results establish a fully integrated photonic platform for atom trapping, paving the way for compact quantum sensors and clocks, which we discuss briefly in Sec.~\ref{sec:app}. Sec.~\ref{sec:conc} concludes. 

\section{\label{sec:loading} Loading the PIC trap}
\subsection{gMOT}
\begin{figure}
    \centering
    \includegraphics[width=1\linewidth]{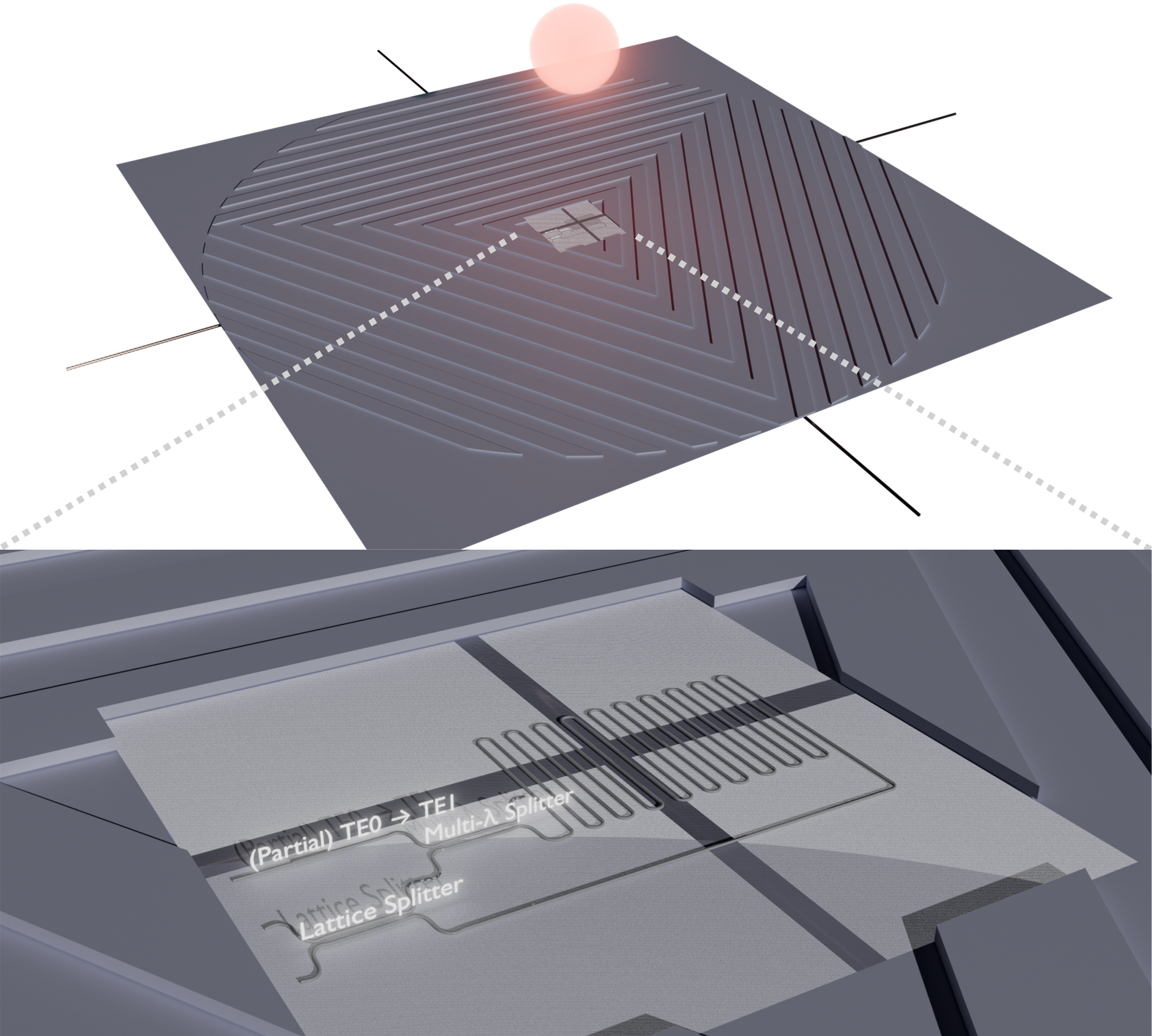}
    \caption{Illustration of the integrated gMOT-PIC system. Atoms are initially trapped using a downwards-incident beam which diffracts to form a MOT cloud. These can then be trapped in the magnetic field generated by current carrying wires (crossed thick wires below the chip), evaporated to BEC, and controllably lowered to the PIC region of the chip (located in the centre of the grating), where the atoms can be trapped in the evanescent fields above the waveguides.}
    \label{fig:BLENDER_pic}
\end{figure}

A widely used technology in miniaturising cold-atom systems is the grating magneto-optical trap. This device consists of $N$ angled regions of 1D phase gratings that diffract a single normally incident beam with efficiency $\eta$, forming an MOT millimetres above the grating's surface \citep{nshii_surface-patterned_2013,goodman_introduction_2017}. The gMOT provides the initial cooling of atoms, before BEC formation and lowering to the PIC trap (Figure \ref{fig:BLENDER_pic}). The condition for balancing such a device is described in Eq. \eqref{eq:gMOT_balancing}, where the diffraction angle $\theta_d$ is defined by the grating equation and the grating's period $(d)$

\begin{equation}
    \sec \theta_d=NR\eta.
    \label{eq:gMOT_balancing}
\end{equation} 
Typical values of a gMOT are $\theta_d=35\degree $ from a period $d=1.4$ µm, efficiency $\eta=0.4$,  and number of regions $N=3$ \citep{cotter_design_2016,nshii_surface-patterned_2013}. For stable trapping at the magnetic field minimum provided by an externally-produced quadrupole field, the incident beam must be appropriately circularly polarised ($\sigma^+ / \sigma^-$) such that each beam drives the correct Zeeman transition relative to the field. In order to produce the required $\pi$ phase change between peaks and troughs, an etch depth of $\lambda/4$ is required (195 nm). An advantage of using this device is that this functionality can be integrated into a complete SiN device using a single-step etch, with the PIC waveguide also of this same height. Through a combination of magneto-optical trapping and optical molasses, cloud temperatures of typically 40 µK - and as low as 3 µK - can be achieved \cite{nshii_surface-patterned_2013}.

\subsection{Wire Trapping}
Following initial gMOT loading, cold atoms can be transferred to a magnetic trap using microwires - lithographically patterned current-carrying conductors on an atom chip that generate strong, localised magnetic fields - and external magnetic fields \cite{calviac_bose-einstein-condensate_2025}. As a result of the Zeeman effect, atoms in low-field-seeking states experience a potential which makes them tend to the minimum of the $|B|$ field $U=-g_Fm_F\mu_B|B|$ (where in low field seeking states $g_Fm_F<0$). Components of the Biot-Savart B-field produced by the $I$ current-carrying wire are cancelled by $B_\textrm{ext}$ at a particular $z_\textrm{trap}\propto \frac {B_\textrm{ext}}{I}$. A variety of wire geometries exist to perform this task, including the Z-trap and the dimple-trap \cite{salim_ultracold_2011}; with the latter envisioned for this system due to providing the most isotropic trap, which improves the re-thermalisation rate of atoms. In order to achieve efficient transfer between traps, the system must be optimised to trap the atoms at the gMOT, transfer adiabatically to the evaporative trap to form a BEC, then transferred to the PIC. Trap parameters for each of these is summarised in Table \ref{tab:gMOT_PIC_transfer}. To ensure that this occurs adiabatically, rapid changes to the spatial profile of the potential should be avoided. This is quantified by the trap frequency $\omega_\textrm{trap}$, which in a harmonic potential is of the form $U=\frac{1}{2}m\omega_\textrm{trap}^2 x^2.$ The vast majority of realistic atom potentials are well-approximated near the trap minimum as harmonic, and this frequency thus gives a measure of the trap geometry.

The adiabatic condition 
\begin{equation}
    \omega_\textrm{trap}^2 \gg\frac{d\omega_\textrm{trap}}{dt}
    \label{eq:wire_adiabatic_condition}
\end{equation}
 must generally be satisfied, although shortcuts to adiabaticity can also be implemented \cite{campo_shortcuts_2013}. 
\begin{table}
    \centering
    \begin{tabular}{|c|c|c|c|}
        \hline
         &  gMOT \cite{nshii_surface-patterned_2013} &Evaporative Trap \citep{weidner_shaken_2018,calviac_bose-einstein-condensate_2025}& PIC Trap\\ \hline
         $z_\textrm{trap}$&  3 mm &3 mm - 250 µm& 200 nm\\
         $\Delta T$&  40 µK &40 µK - 0.3 µK& 1 µK\\
        $\omega_\textrm{trap}$&  $2\pi \times 25$ Hz &$2\pi \times 250$ Hz&$2\pi \times 20$ kHz\\
        \hline
    \end{tabular}
    \caption{Trap parameters for the gMOT, evaporative, and PIC trap to be transferred using the wire trap system. The typical trapping height from the chip is denoted by $z_\textrm{trap}$; $\Delta T=\mu_B\big[U(\textbf{r}_\textrm{trap})-U(\textbf{r}=\infty)\big]$ is the difference in the potential at the trap and far from the trap; and $\omega_\textrm{trap}$ is the trap's frequency.}
    \label{tab:gMOT_PIC_transfer}
\end{table}
Although numerical methods are typically employed to optimise dimple traps in experimental implementations, useful intuition can be gained from the analytic case of a single wire producing a 2D trap. This allows us to estimate the required magnitudes of current $I$ and bias field $B_\textrm{ext}$ to achieve a desired trap height $z_\textrm{trap}$, depth $\Delta T$, and trap frequency $\omega_\textrm{trap}$.

\begin{figure}[H]
    \centering
    \includegraphics[width=1\linewidth]{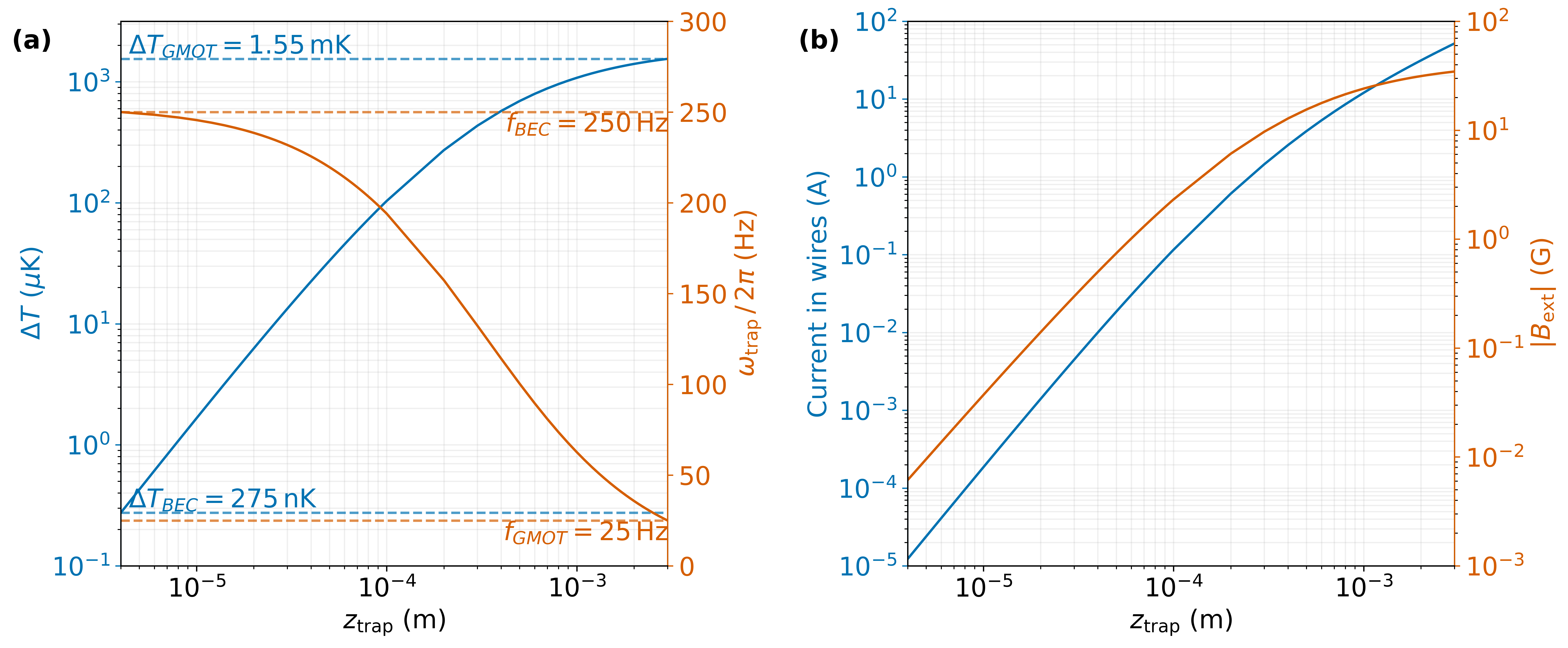}
    \caption{\textbf{(a)} Desired trap parameters $\Delta T$ and $\omega_\textrm{trap}$ against trap height $z_\textrm{trap}$. \textbf{(b)} Required current $I$ and $B_\textrm{ext}$ to produce the 2D trap  with the desired trap parameters against trap height $z_\textrm{trap}$.}
    \label{fig:loading_PIC_trap}
\end{figure}

From the Biot–Savart law, the condition for cancellation of the wire field with the external bias field at the trap centre $\textbf{r}_\textrm{trap}$ is
\begin{equation}
\frac{I}{B_{\text{ext}}} = \frac{2\pi}{\mu_0}\textbf{r}_\textrm{trap}.
    \label{eq:wire_BIOT}
\end{equation}
Expanding the magnetic potential near the trap centre to second order in the transverse coordinate $x$ gives
\begin{equation}
U_x(x,z)=-\mu_\textrm{eff}\bigg[B_\textrm{ext}+\frac{\mu_0I}{2\pi z}-\frac{\mu_0I}{4\pi z^3}x^2 +\mathcal{O}(x^4)\bigg].
    \label{eq:wire_taylor_expansion}
\end{equation}
Here we have taken $\mu_\textrm{eff}=g_F \,m_F \mu_B$. At the trap centre, the constant terms cancel when Eq. \eqref{eq:wire_BIOT} is applied. The quadratic term then reduces to 
\begin{equation}
\Delta U=\frac{\mu_\textrm{eff}\mu_B I}{4 \pi z_\textrm{trap}^3}x^2=\frac{  \mu_B B_\textrm{ext}}{2z_\textrm{trap}^2}x^2.
    \label{eq:wire_taylor_expansion2}
\end{equation}
Comparing this to the harmonic oscillator equation $\Delta U=\frac{1}{2}m\omega^2 x^2$ gives the transverse trap frequency
\begin{equation}
\omega_\textrm{trap}^2= \frac{  \mu_B B_\textrm{ext}}{mz^2_\textrm{trap}}.   
    \label{eq:wire_f}
\end{equation}
This relation makes it clear how the frequency scales with the bias field and trap height. More generally, for a desired trap frequency profile $\omega_d(z)$, the required magnetic field is 
\begin{equation}
B_\textrm{ext}(z)=\frac{mz^2}{\mu_B}\omega_{d}(z)^2,
\label{eq:wire_general_B(z)}
\end{equation}
with the corresponding current $I$ determined by the Biot-Savart condition \eqref{eq:wire_BIOT}. This can be straightforwardly generalised to more complicated three-dimensional traps, where one must add axial bias fields to prevent Majorana spin-flip losses from the field zero at the trap \cite{brink_majorana_2006}. Figure \ref{fig:loading_PIC_trap} illustrates the desired trap parameters (a) and the input parameters (b) from the gMOT $\rightarrow$ evaporative stages, with an analogous procedure governing the evaporative $\rightarrow$ PIC transition. As atoms are lowered closer to the chip surface, the depth at which wires are buried becomes of increasing importance, but this can be used as a free parameter to optimise the potential depth and frequency of the trap for transfer to the PIC trap.   

\subsection{Transferring Between Traps}
Smooth transfer between traps requires controlling the trap frequency to maintain adiabaticity. For a transfer from a low-frequency initial trap at a height and frequency $(h_1,\omega_1)$ - for example the gMOT trap - to a higher frequency trap (evaporative/PIC trap) at $(h_2,\omega_2$), the time dependent trap frequency can be found as a solution to Eq. \eqref{eq:wire_adiabatic_condition}
\begin{equation}
\omega(t)=\frac{\omega_1}{1-\epsilon \omega_1 t}
    \label{eq:wire_w(t)},
\end{equation}
where $\epsilon\ll1$ represents the constant adiabaticity parameter
\begin{equation}
\epsilon=\frac{\omega^2}{\frac{d\omega}{dt}} \Rightarrow\frac{d\omega}{dt}\ll\omega^2.
    \label{eq:wire_epsilon}
\end{equation}
The time taken to adjust the trap's frequency from $\omega_1$ to $\omega_2$ is given by
\begin{equation}
T=\frac{1}{\epsilon} \,\bigg(\frac{1}{\omega_1}-\frac{1}{\omega_2}\bigg)\approx\frac{1}{\epsilon\omega_1}.
    \label{eq:wire_int_sub}
\end{equation}
A repetition frequency $f_\textrm{rep}=\epsilon\omega_1$ can be defined for the system, which is limited at $f_\textrm{rep}\approx157$ Hz for $\epsilon=1$, and slower for more realistic values (ignoring shortcuts to adiabaticity). 
It is also relevant to consider the required $\omega(z)$ in order to calculate the wire trapping parameters. Consider a trap that transfers from $z_1\rightarrow z_2$ in time $T$ at a constant speed,
\begin{equation}
z_\textrm{trap}(t)=z_1+(z_2-z_1)\frac{t}{T}=z_1+\epsilon t(z_2-z_1)\frac{\omega_1\omega_2}{\omega_2 - \omega_1}.
    \label{eq:wire_z(t)}
\end{equation}
We can combine this with Eqs. \eqref{eq:wire_w(t)} and \eqref{eq:wire_int_sub} to find

\begin{equation}
   \omega(z) = \frac{ (h_2 -h_1) \frac{\omega_1\omega_2}{\omega_2-\omega_1} }{ \frac{h_1\omega_1 -h_2\omega_2}{\omega_1-\omega_2} -z} \approx \frac{ (h_2 -h_1)\omega_1}{h_2-z},
\label{eq:wire_w(z)_constant_eps}
\end{equation}
where the approximation is valid for $\omega_2\gg\omega_1$. This mapping allows for a trap frequency to be associated with a given intermediary trap height, which can be used to determine the Biot-Savart currents and bias fields  using Eqs. \eqref{eq:wire_BIOT} and \eqref{eq:wire_general_B(z)} to produce the desired potential at that trap height $z_\textrm{trap}$ (as seen in Figure \ref{fig:loading_PIC_trap}). 

\section{\label{sec:PIC} PIC Trap Design}
\subsection{Design of Vertical Trapping Potentials}
A functional PIC trap relies on red- and blue-detuned modes that evanescently decay out of the waveguide to create a stable potential $\sim200$ nm above the chip's surface. This is similar to schemes produced by fibre traps \citep{morrissey_spectroscopy_2013,vetsch_optical_2010}, with existing literature describing how this could be applied in PICs using air-clad suspended SiO$_2$ waveguides \citep{ovchinnikov_two-mode_2025,ovchinnikov_towards_2020}. In this section, we will discuss a platform for atom trapping using buried rectangular strip SiN waveguides in SiO$_2$ cladding. SiN represents a good platform choice due to its low loss transmission at visible/near-IR wavelengths of atomic transitions, its tolerance for small bend radii/high powers, and its ability to integrate laser sources on chip \citep{blumenthal_silicon_2018,liu_large_2025}. The platform has an increased fabrication maturity which has enabled its widespread use for commercial, research, and quantum technological applications \citep{alexander_et_al_psiquantum_team_manufacturable_2025,blumenthal_enabling_2024}.

\begin{figure*}
    \centering
    \includegraphics[width=1\linewidth]{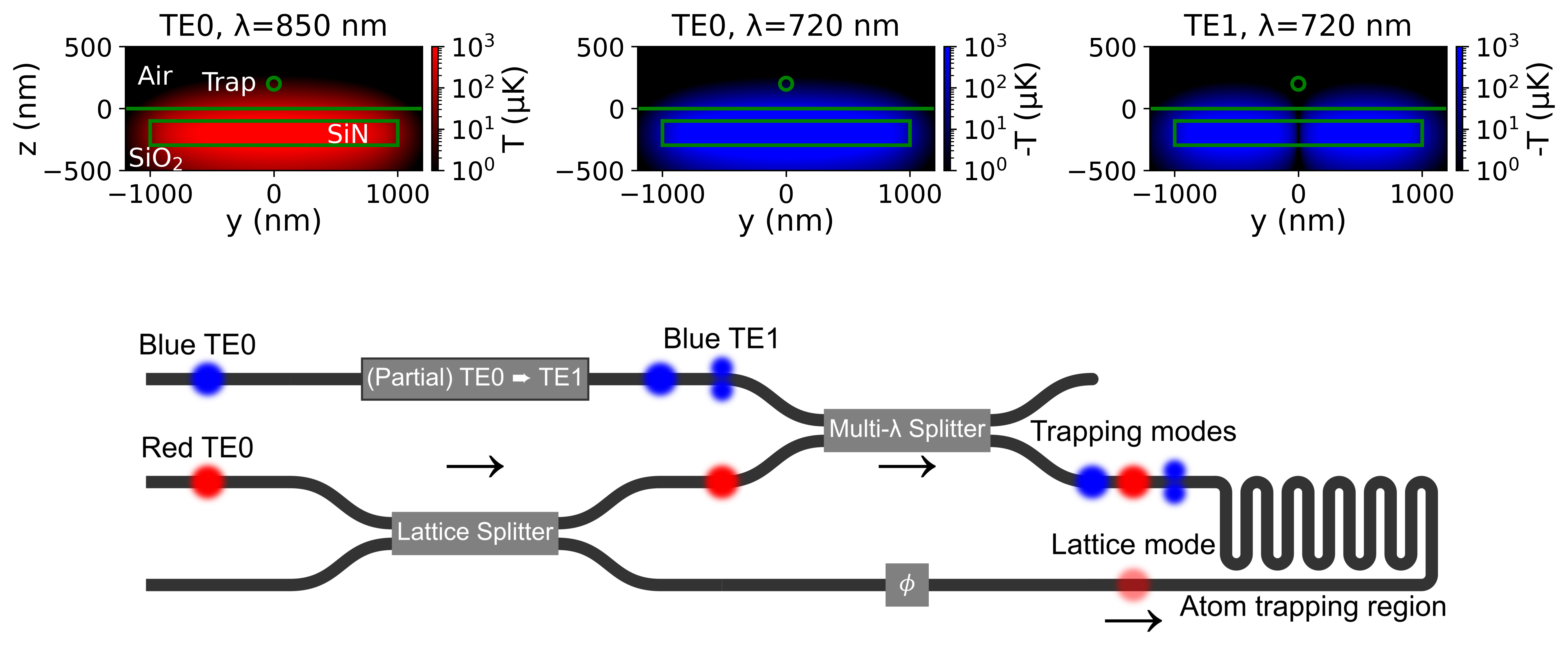}
    \caption{\textbf{(Top)} Red TE0, blue TE0, and red TE1 spatial potentials from a SiN waveguide in buried silica, expressed in effective temperature $T=\frac{2}{3}\frac{U}{k_B}$ which is proportional to $|E_y|^2$. \textbf{(Bottom)} Schematic of the combination of red, blue, lateral, and lattice modes to produce copropagation in a single snaking waveguide in the atom trapping region.}
    \label{fig:PIC_trap_modes}
\end{figure*}

The waveguide geometry consists of a 2 µm $\times$ 195 nm Si$_3$N$_4$ rectangular strip embedded in SiO$_2$, with a 100 nm top-oxide cladding above the upper surface of the core which forms the surface of the chip at $z=0$. All modes considered in this work are TE-polarized, with the electric field oriented along the $\hat{y}$ direction. The potential $U_n$ from a given mode $n$, with angular frequency $w_n$ and intensity $I_n(\textbf{r})$, from the $D_j$ transitions with natural linewidths $\Gamma_j$ of $^{87}$Rb is given by

\begin{equation}
     U_n(\omega_n,\textbf{r})=\pi c^2  \sum_{j=1,2}\frac{j\space\Gamma_j}{\omega_j^2(\omega_n^2 -\omega_j^2) }  I_n(\textbf{r}).
     \label{eq:PIC_w_T_relation}
\end{equation}
The $n$ modes used are plotted in Figure \ref{fig:PIC_trap_modes}, where the potential is expressed in effective temperature $T=\frac{2}{3}\frac{U}{k_B}$ and is proportional to $|E_y|^2$. This can be attractive or repulsive depending on the mode's red or blue detuning from the $j=\{1,2\}$ transitions at $\omega_j$ \cite{grimm_optical_2000}. Light can be coupled into the PIC waveguides via traditional grating-based or edge-coupling methods~\cite{Marchetti2019} or generated via on-chip laser sources~\cite{Moody2025,heim_hybrid_2025,liu_large_2025,gallacher_integrated_2019}. In order to produce a barrier potential for the atoms above the chip's surface, a blue mode must be incorporated such that $U_\textrm{blue}(z=0)>U_\textrm{red}(z=0)$, so that the potential is repulsive. Fortunately, the exponential decay rate linearly increases with $\lambda$ in the visible/near-IR range as the mode reddens, with $d_{red}<d_{blue}$, with the evanescent decay length from the chip's surface taking the form $d(\lambda)=0.0718\lambda-11.88$ (found via numerical simulation). This means that at a height of $z_\mathrm{trap}\approx200$ ~nm from the chip's surface, the repulsive potential's magnitude dips below that of the attractive potential to form a trap, before evanescently decaying as $z\to\infty$. The total potential from the red and blue modes is given by
\begin{equation}
   U(z)=\sum _n U_n(z=0)e^{-z/d_n} + U_\textrm{CP}.
   \label{eq:PIC_combined_potential}
\end{equation}
The Casimir-Polder (CP) force describes the attractive potential in the regime of atoms interacting with a dielectric surface 0.1-1 µm away going roughly as $U_\textrm{CP} \sim1/z^4$ for a planar surface \citep{antezza_effect_2004,klimchitskaya_casimir_2009,stern_simulations_2011}. An effective approximation for the CP force for $^{87}$Rb that incorporates the behaviour at $z\sim0.1$µm  is given by Eq.\eqref{eq:PIC_CP}, where $C_3=2\pi\hbar \space\times860$ Hz µm$^3$ and $\lambdabar=710  /(2\pi)$nm for SiO$_2$ \cite{thompson_coupling_2013}:

\begin{equation}
U_\textrm{CP}(z)=-\frac{C_{3}\lambdabar}{z^3(z+\lambdabar )}.
    \label{eq:PIC_CP}
\end{equation}
This potential $U_\textrm{CP}\to -\infty$ as $z\to0$ meaning $U_\textrm{CP}(z=0)>U_\textrm{red}(z=0)$ and no infinite barrier can be formed to the surface. However, a barrier significantly higher than the trap depth can be formed, and the CP potential can be incorporated to improve the trapping parameters. An advantage that a uniform 100 nm thick top oxide coating of SiO$_2$ provides is that there is no $x-y$ dependence in Eq.\eqref{eq:PIC_CP}. We will begin by exclusively considering TE0 modes for a trap along the central $z$ axis at $y=0$. This is because $U_\textrm{TE1}(y=0)=0$ is a condition of solving Maxwell's equations for a TE1 waveguide mode, and hence doesn't contribute along this axis. Lateral trapping using a higher order mode is considered in Section \ref{sec:lateral_trapping}. The position of the trap and the barrier formed by the potentials in Eq. \ref{eq:PIC_combined_potential} are at $\frac{dU(z)}{dz}=0$, with their second derivatives at $z$ being positive and negative, respectively. 

\begin{figure}[H]
    \centering
    \includegraphics[width=1\linewidth]{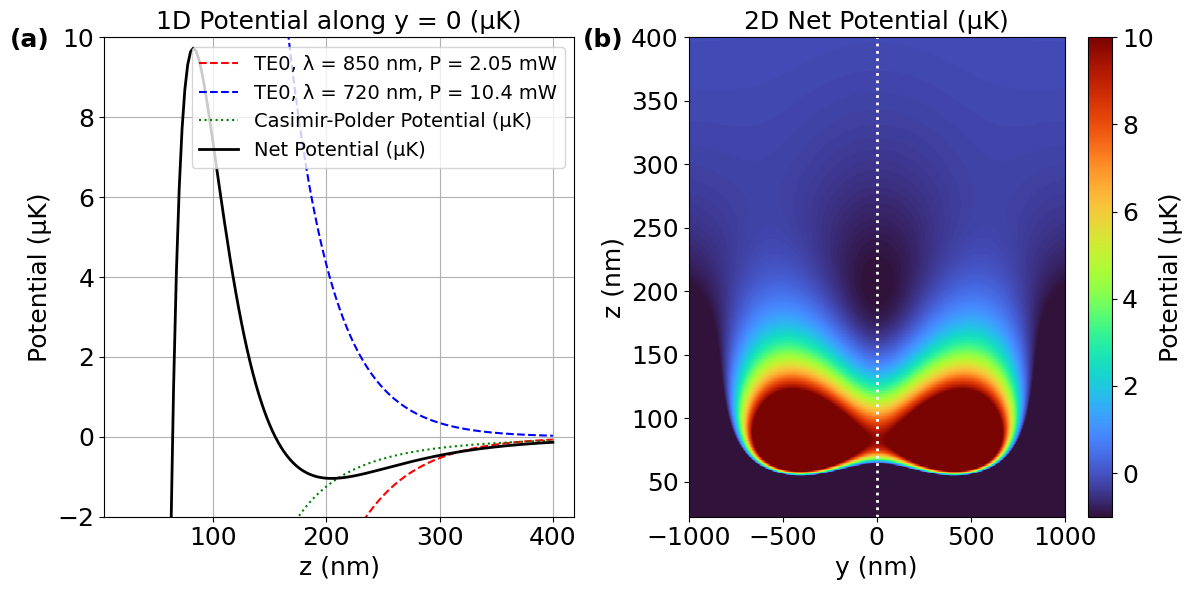}
    \caption{\textbf{(a)} Simulation of the potential of the atoms expressed in µK along the central $y=0$ axis. A clear peak and trough is formed for atom trapping 205 nm above the waveguide surface. \textbf{(b)} 2D simulation of the trap including a lateral trapping TE1 mode for two-dimensional confinement ($y=0$ line from (a) overlaid).}
    \label{fig:PIC_trap_profile}
\end{figure}

\begin{figure*}
    \centering
    \includegraphics[width=1\linewidth]{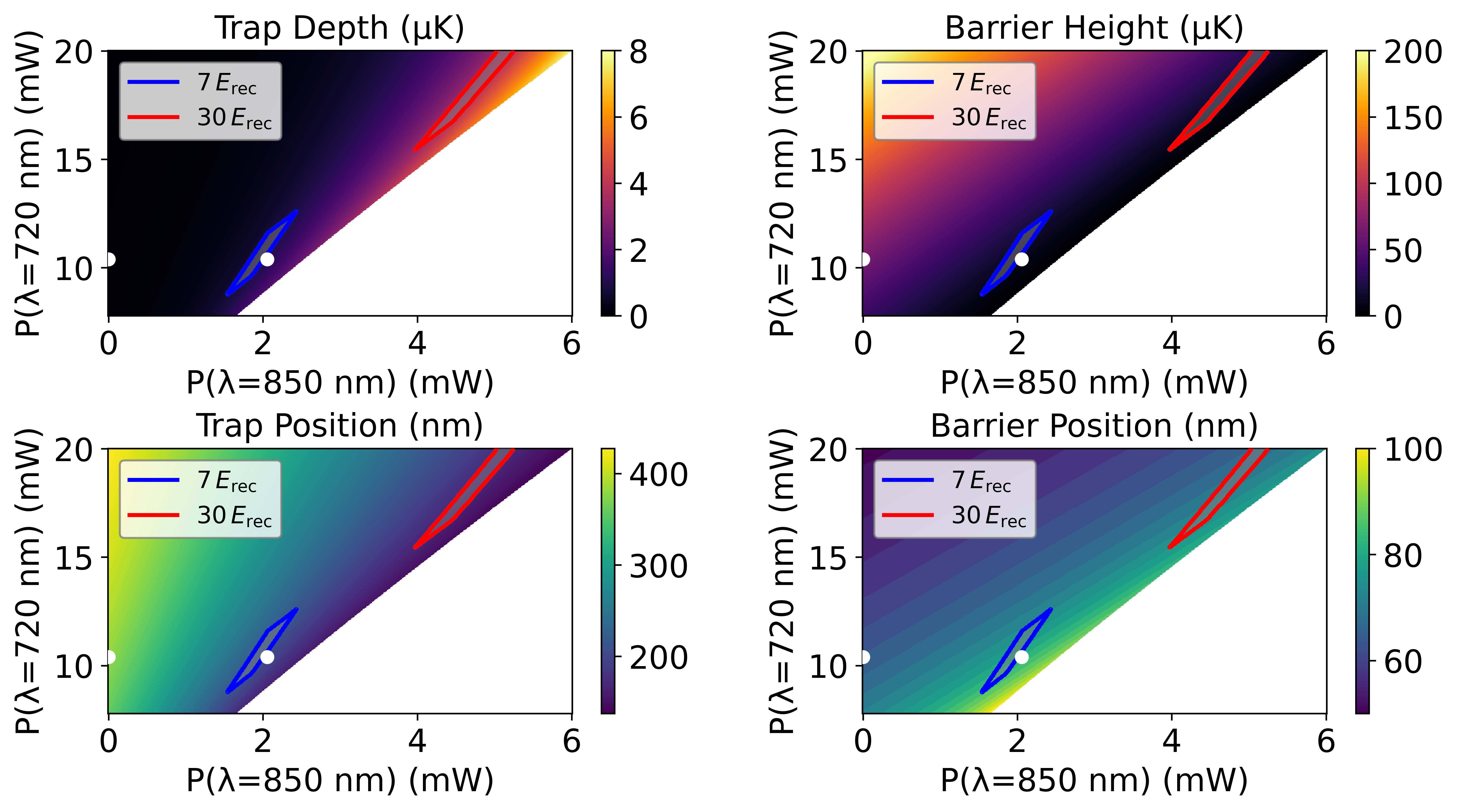}
    \caption{Effect of sweeping blue and red laser power on the trap/barrier's potential and position along the $z$ axis. Regions are marked where $\Delta T_\textrm{trap}\approx\{7,30\}\, E_\textrm{rec}$. The minimum viable trap with the $7E_\textrm{rec}$ criterion used predominantly in this paper is at $P_{r}$ = 1.65 mW and $P_{b}$ = 9 mW. The rightmost white dot represents the trap parameters plotted in Figure \ref{fig:PIC_trap_profile} and the leftmost dot is the potential at saddle points in the optical lattice formed (see Section \ref{sec:standing_wave}).}
    \label{fig:PIC_trap_parameters}
\end{figure*}

Equation \eqref{eq:PIC_combined_potential} is a transcendental equation that cannot be solved analytically for $z$, but numerically we can calculate what values of $U_{r,b}$ are required to produce a trap of depth $\Delta T_\textrm{trap}$ and a barrier $\Delta T_\textrm{barrier}$ (Table \ref{tab:PIC_parameters}). Figure \ref{fig:PIC_trap_profile}(a) shows the shape of this potential along the centre of the waveguide $y=0$, and a stable trap of $\Delta T=1$ µK is formed at $z=205$ nm. Figure \ref{fig:PIC_trap_parameters} describes the mode powers incident down the trapping waveguide and their effect on the barrier/trap's potential height/depth. We adopt acceptability criteria of trap depth $7(\pm 1)E_\textrm{rec}$ and $30(\pm 3)E_\textrm{rec}$, together with a barrier potential of $8$ µK $<\Delta T_\textrm{barrier}<20$ µK. These regions are highlighted to illustrate this system's tunability and robustness to fluctuations in laser power. 
\begin{table}
    \centering
    \begin{tabular}{|c|c|c|}
    \hline
         Desired Parameters&$\Delta T_\textrm{trap}$ &-1 µK   \\
          &$\Delta T_\textrm{barrier}$&10 µK   \\
         \hline
         &$\lambda_r$ &850 nm   \\ 
  Wavelength Choices&$\lambda_b$ &720 nm\\
  &$d_r$&48.9 nm\\
  &$d_b$&39.7 nm\\
  \hline
 & $T_r (z=0)$&-244 µK\\
 & $T_b (z=0)$&667 µK\\
 Derived Parameters& $P_r$&2.05 mW\\
 & $P_b$&10.4 mW\\
 & $z_\textrm{trap}$&205 nm\\
 & $z_\textrm{barrier}$&80 nm\\
  \hline
    \end{tabular}
    \caption{Trap and barrier choice parameters and their influence on the mode powers required, as well as the $z$ positions that these features give rise to. Atom potentials have been converted to effective temperature $T=\frac{2}{3}\frac{U}{k_B}$, explaining the negative signs.}
    \label{tab:PIC_parameters}
\end{table}

\subsection{Three-Dimensional Trapping}\label{sec:lateral_trapping}
\subsubsection{Lateral Trapping}

Once $z_\textrm{trap}$ for our system has been calculated, a blue-detuned TE1 mode can be introduced to produce lateral trapping without affecting the trapping along the central axis, as $U_{TE1}(y=0)=0$. No laterally stable trap is formed without including such a mode. The effect of the introduction of a TE1 mode is seen in Figure \ref{fig:PIC_trap_profile}(b). Two important considerations can be made to aid the design of the TE1 mode. Firstly, the mode profiles of the red and blue TE0 modes are approximately equal above the central portion of the waveguide. Secondly, the height of the trap is formed where $U_{r,TE0}(z)=-U_{b,TE0}(z)$, meaning at the modes' potentials are balanced at this point, and the potential's magnitude is determined predominantly by the CP force, i.e., $U(y,z_\textrm{trap})=U_\textrm{CP}$. This is true across the waveguide's width due to the planar top oxide surface producing a laterally constant CP force, and the TE1 modes' matched profiles. Therefore to produce a lateral trapping potential, the TE1 mode's magnitude should obey

\begin{equation}
U_{b,TE1}(\pm y_\textrm{max},z_\textrm{trap})=-U_\textrm{CP}(z_\textrm{trap})+\Delta U_{\textrm{y offset}},
    \label{eq:PIC_TE1_CP}
\end{equation}
where $\Delta U_{\textrm{y offset}}$ is a desired increase to the potential above zero (which is omitted in this instance).  This potential can be readily converted to calculate the required total power in the waveguide (4 mW). The positions of the TE1 mode maxima are at $y_\textrm{max}=\pm 0.2465w_\textrm{wg}$, which only differ slightly from $w_\textrm{wg}/4$ due to the TE0 modes not being perfectly balanced at $z_\textrm{trap}$.  

\subsubsection{Lattice Trapping}\label{sec:standing_wave}
\begin{figure}
    \centering
    \includegraphics[width=1\linewidth]{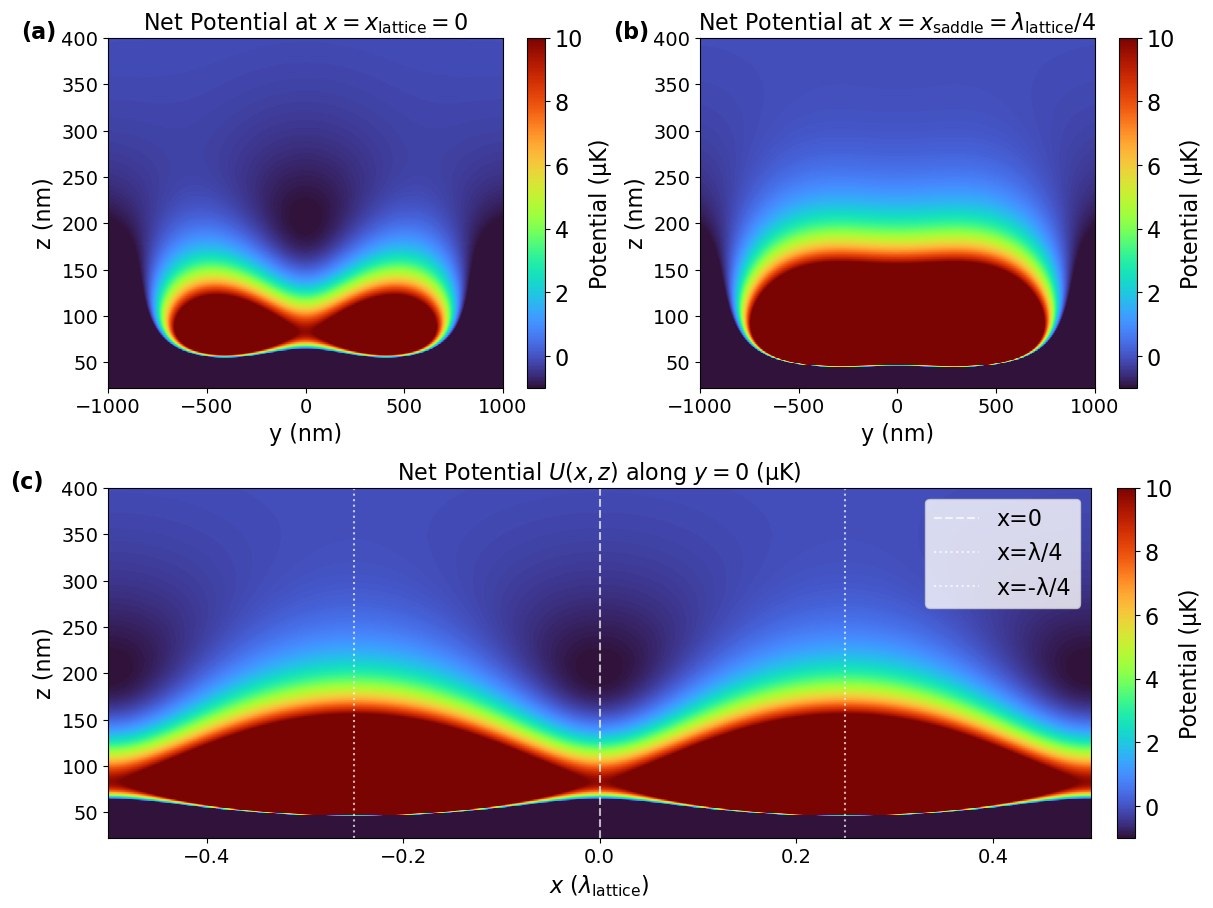}
   \caption{\textbf{(a)} Previous potential and potential at $x_\textrm{lattice}=2m \cdot\frac{\lambda_\textrm{lattice}}{4}$ for $m \in \mathbb{Z}$. \textbf{(b)} Potential at $x_\textrm{saddle}=(2m-1) \cdot\frac{\lambda_\textrm{lattice}}{4}$, where $\Delta T\approx0$. Some lateral confinement exists but atoms will tend to the potential minima at lattice sites. \textbf{(c)} Potential along $\hat{x}$ against height $z$ along the plane $y=0$.}
    \label{fig:PIC_saddle}
\end{figure}

This system is compatible with forming an optical lattice potential along the propagation direction of the waveguide $\hat x$. This can be formed with either the red or blue TE0 mode also being incident in the opposing direction of the waveguide. Examining Figure \ref{fig:PIC_trap_parameters}, we want to produce our previous trap at lattice sites $x_\textrm{lattice}=2m \cdot\frac{\lambda_\textrm{lattice}}{4}$ for $m \in \mathbb{Z}$, and at the lattice saddle points $x_\textrm{saddle}=(2m-1) \cdot\frac{\lambda_\textrm{lattice}}{4}$, there is a weaker trap ($\Delta T_\textrm{saddle}\approx0$ K); with the period of the trap lattice being $\frac{\lambda_\textrm{lattice}}{2}$. Such saddle points can be achieved either by having saddle sites with decreased red intensity or increased blue intensity. In this case, we will consider a red-detuned lattice as it involves decreasing the light intensity at saddle points (reducing scattering), and it has a marginally larger period, and hence trap volume. 
Consider making an 850 nm lattice of visibility $V=\frac{P_\mathrm{lattice}-P_\mathrm{saddle}}{P_\mathrm{lattice}+P_\mathrm{saddle}}=1$. The effective mode power at $x_\textrm{saddle}$ is $P_{r,TE0}(x_\textrm{saddle})=$ 0 mW, with the desired power at the lattice points $P_{r,TE0}(x_\textrm{lattice})=$ 2.05 mW as before, which would result in a lattice depth of $\sim 7 E_\textrm{rec}$. The total power required in the red TE0 modes is the mean of these powers $\overline{P_{r,TE0}} =\frac{P_\mathrm{lattice}+P_\mathrm{saddle}}{2}=1.025$ mW. To achieve these intensities, the incident $\overline{P_{r,TE0}}$  must be split equally by a ratio $r_P=0.5$ in the forward and backward intensities. More generally, the splitting ratio $r_P$ required can be calculated according to
\begin{equation}
r_P=\frac{1+\sqrt{1-V^2}}{2}.
    \label{eq:PIC_alpha}
\end{equation}
The design criterion of this splitting ratio being $r_P\approx0.5$ is quite relaxed - with $r_P=0.4$ still providing 99\% of the intensity desired at $x_\textrm{lattice}$. This results in a potential of the form seen in Figure \ref{fig:PIC_saddle}, which still provides confinement in the $\hat{y},\hat{z}$ directions, but due to the attractive nature of the lattice trap antinodes, the atoms will tend towards the lattice sites.

Note that in this work, we will also consider a lattice depth of $30E_\textrm{rec}$. These two depths, $7E_\textrm{rec}$ and $30E_\textrm{rec}$, allow us to demonstrate points where the atoms are deep within the delocalized superfluid or localized insulating regime, respectively~\cite{Sapiro2009}, both of which are useful for quantum technological applications, as we describe in Sec.~\ref{sec:app}. 

\subsection{Trap Properties}
With full three-dimensional spatial confinement achieved, trap frequencies along all dimensions can be calculated, with values $\sim10$ kHz. Trap frequencies can be readily modified in $\hat{x}$ and $\hat{y}$ dimensions by altering $\lambda_{\textrm{lattice}}$ and $w_\textrm{wg}$, respectively. The trap frequency at the lattice and saddle points with our previous parameters are found in Table \ref{tab:PIC_frequency}. 

\begin{table}
    \centering
    \begin{tabular}{|l|ccc|c|c|}
\hline
 &$f_x$ (kHz) &$f_y$ (kHz) &$f_z$ (kHz) & $z_\mathrm{trap}$ (nm)&$\Delta T$ (µK)\\
 \hline
          $\textbf{r}_\textrm{lattice}$&16.9&9.5 &38.9 & 205&1\\
          
  $\textbf{r}_\textrm{saddle}$&-6.25& 0.6&7.1& 380&0.07\\
  \hline
    \end{tabular}
    \caption{Numerical calculations of the trap frequencies in all dimensions, positions, and depth at $\textbf{r}_\textrm{lattice}$ and  $\textbf{r}_\textrm{saddle}$. These locations are independently calculated, as $z_\textrm{lattice}\neq z_\textrm{saddle}$,  with $z_\textrm{saddle}=380$ nm. A negative value of $f_x(\textbf{r}_\textrm{saddle})$ illustrates the saddle potential's steepness.}
    \label{tab:PIC_frequency}
\end{table}

The photon scattering rate for alkali atoms such as $^{87}$Rb, considering contributions from both the D1 and D2 transitions, can be expressed as
\begin{equation}
  \Gamma_{\mathrm{sc}}(\mathbf{r})=\frac{\pi c^2 }{2 \hbar \omega_0^3}\left(\frac{2\Gamma_2^2}{\Delta_{2}^2}+\frac{\Gamma_1^2}{\Delta_{1}^2}\right) I(\mathbf{r}).
\label{eq:PIC_Scattering_rate}
\end{equation}
The detuning is defined as $\Delta_j\equiv\omega_n-\omega_j$, and the equation applies  in the far-detuned approximation - where $\frac{\Delta_j}{\Gamma_j}\gg1$, which in this system's case is $\sim10^6$ \citep{grimm_optical_2000,weidner_shaken_2018}. When combined with our earlier expression for the potential in Eq.\eqref{eq:PIC_w_T_relation}, we can ascertain the scattering rate $\Gamma_\textrm{sc}$ as a function of the potential $U(\textbf{r})$,
\begin{equation}
\Gamma_{{sc}}=\frac{\frac{\Gamma_1^2}{\Delta_1^2}+2 \frac{\Gamma_2^2}{\Delta_2^2}}{\frac{\Gamma_1}{\Delta_1}+2 \frac{\Gamma_2}{\Delta_2}} \cdot \frac{U(\textbf{r})}{\hbar}.
    \label{eq:PIC_scattering}
\end{equation}

We can then use the potential at $z_\textrm{trap}$ to calculate  $\Gamma_{sc,r}(\textbf{r}_\textrm{trap})=0.16  $ s$^{-1}$ and $\Gamma_{sc,b}(\textbf{r}_\textrm{trap})$ = 0.20 s$^{-1}$, which gives $\sum_{\lambda} \Gamma_{sc,\lambda}=0.36$ s$^{-1}$. This results in a heating rate of $\frac{dT}{dt}=\frac{2E_\textrm{rec}}{3k_B}\sum_{\lambda} \Gamma_{sc,\lambda}\approx130$ nK/s. The lifetime of the trap $t_\textrm{life}$ can also be calculated using the recoil temperature of the transition $T_\textrm{rec}=\frac{\hbar k_j^2}{mk_B}$,
\begin{equation}
t_\textrm{life}=T_\textrm{trap}/(T_\textrm{rec} \space\Gamma_\textrm{sc})\approx7.7\textrm{s}.
    \label{eq:PIC_t_life}
\end{equation}
Importantly, this is much longer-lived than the rate at which the trap can be loaded ($f_\textrm{rep}=\epsilon \omega_\textrm{gMOT}\sim\epsilon157$ Hz), and even lower scattering rates can be achieved through further detuning, as $\Gamma_\textrm{sc}\propto\frac{I}{\Delta^2}$. This will however increase the amount of power required within the chip, as $U \propto \frac{I}{\Delta}$, but this trade-off scales favourably, especially considering that a significantly higher intensity can be handled by the SiN chip than is currently incident, with $P>100$ mW with a smaller mode effective area reported to be achievable \cite{tien_ultra-low_2010}. Further detuning would lead to lower trap heights (due to a larger gap between decay rates $d_r$ and $d_b$), and other atomic transitions may need to be considered \cite{ovchinnikov_two-mode_2025}.

\subsection{Component Design}\label{sec: Component Design}
Useful PIC components for atom trapping using SiN exist within the literature, including TE0 to TE1 converters, low loss splitters, phase modulators, and polarisation converters \citep{mcgilligan_micro-fabricated_2022,gallacher_silicon_2022,januszewicz_chip-scale_2025}. These are designed for use on resonance at a single wavelength (780 nm), so challenges remain in designing components for simultaneously addressing modes that are red- and blue-detuned from resonance. 
For the lattice-splitter that provides the red trapping and lattice modes (section \ref{sec:standing_wave}), a near 100\% output is required in the trapping output, with minimal diversion towards the lattice - this hence requires careful engineering to ensure the power balance is acceptable. This is simplified by the fact that the splitter is operated at a single wavelength. The lattice mode needs to be phase modulated to provide transport of atoms for applications such as shaken lattice interferometry~\cite{weidner_experimental_2018} or clocks~\cite{Katori,Schmidt}, and this can be achieved using on-chip phase modulators \citep{alexander_nanophotonic_2018,januszewicz_chip-scale_2025}. To transport the atoms along the waveguide, modulating the lattice mode by $2\pi$ moves the lattice by a distance $\frac{\lambda_\textrm{lattice}}{2}=425$ nm in this instance. A modulation frequency of $f_\textrm{mod }=$ 1 kHz results in a lattice velocity $v_\textrm{lattice}=\frac{\lambda_\textrm{lattice}}{2f_\textrm{mod}}=0.42$ mm/s. For inertial sensing applications (see Section \ref{sec:intertial sensing}),  $f_\textrm{mod}\approx 0.5$ MHz are required \citep{kendell_deterministic_2024}, with the possibility of transitioning to LiNbO$_3$ hybrid integration or waveguides, which have intrinsically high electro-optic coefficients whilst being low loss at $\lambda\sim780$ nm.

PIC mode converters are also required to construct the trap. The total blue mode intensity (TE0 + TE1) is 13.4mW, with about 78\% and 22\% needed in each mode respectively. Asymmetrical single-waveguides and directional coupler mode converters are able to achieve 95\% mode-conversion efficiency \citep{gallacher_silicon_2022,sulway_novel_2022}; meaning this fraction is achievable to engineer, as full mode conversion is not required. 

\begin{figure}
    \centering
    \includegraphics[width=1\linewidth]{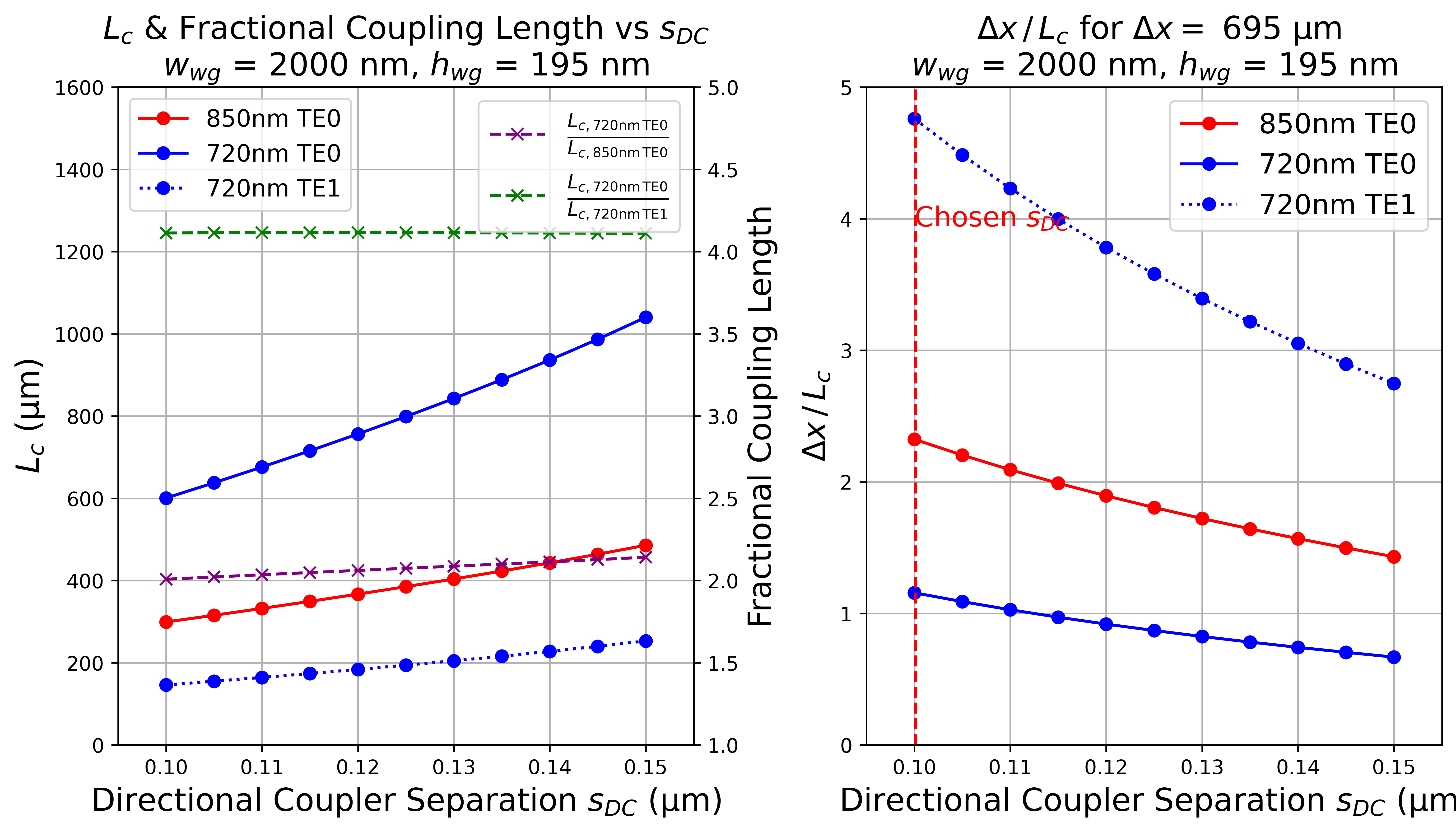}
    \caption{\textbf{(a)} Coupling lengths $L_c$ for red TE0, blue TE0, and blue TE1 modes against DC separation $s_{DC}$. Fractional comparisons to the blue TE0 mode are also plotted, where advantageously $L_C(\lambda=720\text{nm, TE0})\approx 2L_C(\lambda=\text{850nm, TE0})$, but regrettably $L_C(\lambda=\text{720nm, TE0})\approx 4L_C(\lambda=\text{720nm, TE1})$. \textbf{(b)} An interaction length $\Delta x=695$nm is chosen, and the effect on how many $L_c$ each mode experiences is plotted. An $s_{DC}=100$nm is chosen, which means each mode goes through $\{1.1, 2.2,4.8\}\,L_c$ respectively. This yields a total loss $L=0.51\text{dB}$.}
    \label{fig:L_c}
\end{figure}

The remaining challenge is to combine these multi-$\lambda$ modes propagating in separate waveguides into a single waveguide with phase preservation. The design of directional couplers and multi-mode inteferometers  is built on the fact that the propagation of the even/odd supermodes (where the electric field oscillates in/out of phase between each arm) in the interaction region have different effective refractive indices $n_\mathrm{even}$ and $n_\textrm{odd}$. By considering the superposition of the propagation of these modes in the interaction region, an equation for the coupling length $L_c$ can be produced, which is the length at which the power is fully transferred into the other waveguide mode
\begin{equation}
L_c=\frac{\lambda}{2(n_\textrm{even}-n_\textrm{odd})}.
    \label{eq:PIC_L_c}
\end{equation}
For blue and red modes in inputs 1 and 2 (top and bottom of the Multi-$\lambda$ splitter in Figure \ref{fig:PIC_trap_modes}, respectively), both fully combined into output 2, this would require a design of a beam splitting device such that

\begin{equation}
L=(2m-1)L_{c,\,b}=2m \, L_{c,\,r} \quad \text{for } m\in\mathbb{Z}^+.
    \label{eq:PIC_L_C_2_colour}
\end{equation}
This coupling length can be adjusted using the directional coupler gap $s_{DC}$ or alternatively the width of the multi-mode inteferometer, $w_\textrm{MMI}$. A directional coupler is simulated using the chosen waveguide dimensions $w_\textrm{wg}$ and $h_\textrm{wg}$ in Figure \ref{fig:L_c}. A separation $s_{DC}=100$ nm and an interaction length of $\Delta x_{DC}=583$ µm is found to provide full transfer of TE0 720nm to the opposite output $(m=1)$ and self imaging of the 850nm TE0 mode into the same output $(m=2)$. However, in this geometry only 3\% of the 720 nm TE1 mode is converted into the correct mode as $\frac{\Delta x_{DC}}{L_c(TE1,720nm)}=4.11$ and $P_2=\sin^2 \big(\frac{\pi}{2}\frac{\Delta x_{DC}}{L_c(TE1,720nm)}\big)=3.3\%=-15\,\text{dB}$ loss in this mode. An alternative approach arises by considering the required output powers for each of the modes to find an optimised $\Delta x_{DC}$ and $s_{DC}$ that minimises the total power wasted
\begin{equation}
\sum_n P_{n,\text{in}}(\Delta x_{DC},s_{DC})
=\sum_{n}\frac{P_{n,\text{out}}}{\sin^2\!\Big(\dfrac{\pi\Delta x}{2L_{C,n}(s_{DC})}+\phi_n\Big)}.
    \label{eq:PIC_Power_min}
\end{equation}
$P_{in}$ is minimised over the $n$ modes, where $\phi_{TE1,850nm}=\frac{\pi}{2}$ and is zero otherwise (distinguishing between switching output or preserving output). The loss is calculated by computing $ L=\sum_n \big[P_{n,\text{in}}(\Delta x,s_{DC})/P_{n,\text{out}}\big]$. A design with $\Delta x_{DC}=695$ µm and $s_{DC}=100$ nm  provides the required power at the output arm of the directional coupler with a loss of only $L=0.51\text{ dB}$ of loss over all the input modes. Other solutions exist at higher separations with only slightly higher loss $L<1\textrm{ dB}$ if manufacturing tolerances need to be considered, although in all cases interaction lengths $\Delta x>400$ µm are required in order to get reasonable modulation. 

\section{\label{sec:banana}Potential System Failure Modes}
\subsection{Stray Charges}
\begin{figure}
    \centering
    \includegraphics[width=1\linewidth]{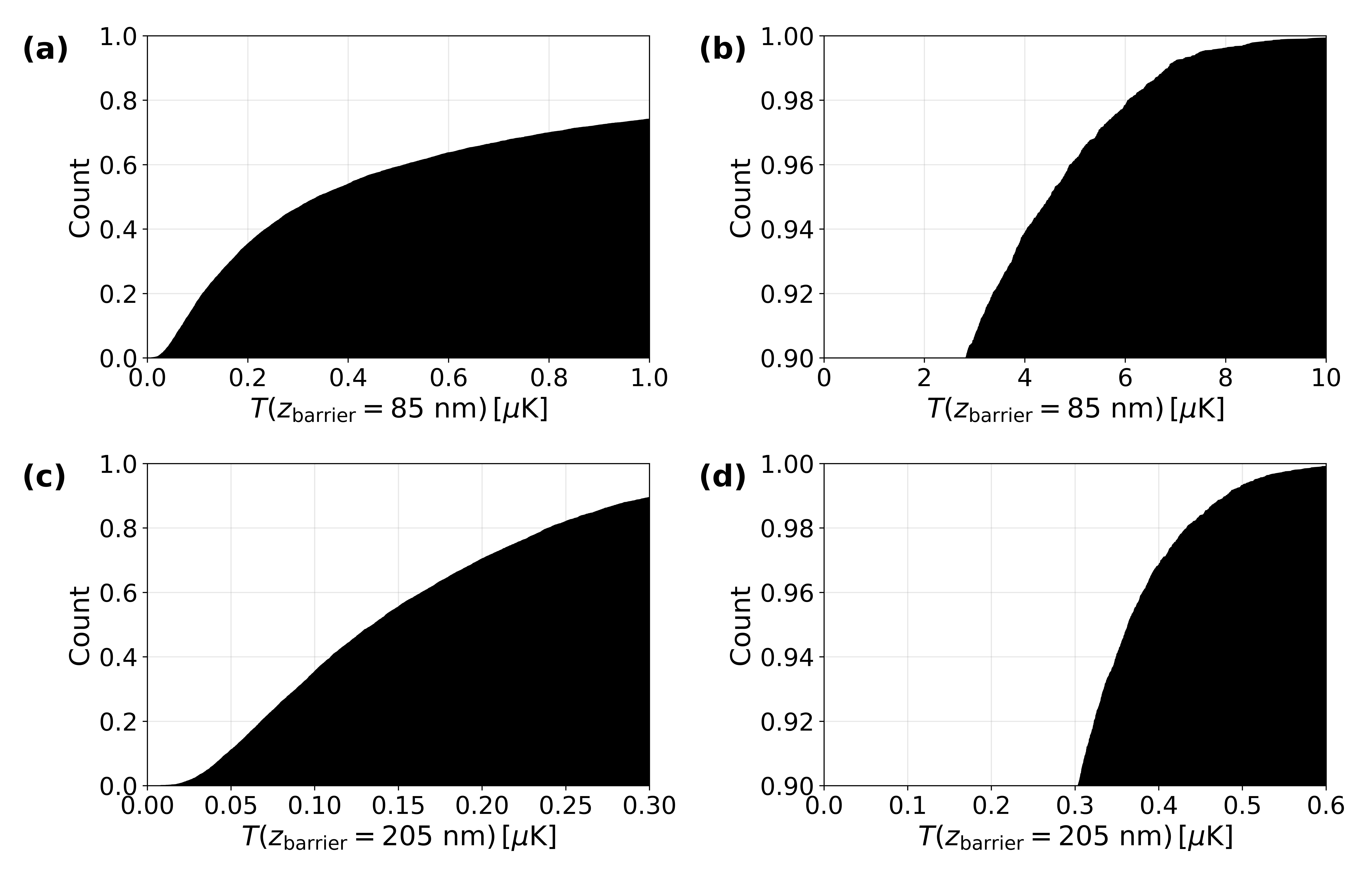}
    \caption{Histogram showing the effective temperature shift induced in the vertical direction by stray charges over $10^5$ instances. Charges are randomly distributed across the surface in lab conditions (charge density of $\sigma=10^{-6}$C/m$^2$). This is considered at the barrier \textbf{(a,b)}; trap \textbf{(c,d)}; over a large range \textbf{(a,c)}; and the worst 10\% of instances \textbf{(b,d)}.}
    \label{fig:charge_histogram}
\end{figure}

The effect of stray charges and imperfections on the surface and within the chip is a potential source of heating and loss within the trap \cite{keil_fifteen_2016}. Due to these, atoms in low-lying or ground states experience a potential $U\propto \frac{1}{r^4}$ - Eq.\eqref{eq:Charge_single}) - which is much weaker than ions ($U\propto \frac{1}{r}$) or Rydberg atoms where the principal quantum number $n\approx50$, and $\alpha\propto n^7$ (compared to $n=5$ for the $^{87}$Rb ground state) \citep{davtyan_controlling_2018,teller_heating_2021,yerokhin_electric_2016}. Regardless, due to the comparably short distance between the chip and atoms, alkali adsorbates such as $F^-$ and $OH^-$ can be present due to clean-room techniques, as well as charges from humid air, causing patch DC potentials \cite{keil_fifteen_2016}. A uniformly charged surface only applies a global shift in the potential proportional to the surface charge density squared $(\sigma^2)$, which would not be too problematic. However, using the potential induced by a point charge and then distributing these randomly yields a positionally dependent force arises from the atom's polarisability $\alpha$, 
\begin{equation}
U=-\frac{1}{2}\alpha E^2=-\frac{1}{2}\frac{q^2\alpha}{(4\pi \epsilon_0)^2 r^4}.
    \label{eq:Charge_single}
\end{equation}
The charge density can be categorised into three major regimes - highly contaminated surfaces with $\sigma=10^{-4}$ C/m$^2$, which is typical in humid air; laboratory conditions with $\sigma=10^{-6}$ C/m$^2$; and ultra-clean surfaces obtained with vacuum/plasma annealing and $N_2$/vacuum storage with $\sigma=10^{-8}$ C/m$^2$ \citep{chan_adsorbate_2014,luo_numerical_2025}. The median effective temperature shifts induced by these charges at $z_\textrm{trap}$ and $z_\textrm{barrier}$ are listed in Table \ref{tab:charge}. Figure \ref{fig:charge_histogram} shows the distribution obtained from Monte Carlo simulations under laboratory conditions. Contributions across the $x-y$ plane are much weaker due their effects counteracting, and the only significant asymmetry existing in the $z$ plane. At $10^{-8}$C/m$^2$, more than $99$\% of samples were below 100 nK at $z_\textrm{trap}$, illustrating that these conditions are sufficient to yield a stray charge profile that doesn't appreciably affect the trapping of atoms.
\begin{table}
    \centering
    \begin{tabular}{|c|cc|}
    \hline
         $\sigma$ (C/m$^2$)&  median$[T_\sigma (z_\textrm{trap})]$&  median$[T_\sigma (z_\textrm{barrier})]$\\
         \hline
         $10^{-4}$& 10.6 µK & 88 µK  \\
         $10^{-6}$&0.14 µK&  0.5 µK\\
         $10^{-8}$&  $<$1 nK&  $<$1 nK\\
         \hline
    \end{tabular}
    \caption{Median effective temperature shifts due to surface charges in the vertical direction. These values indicate that charge densities below $10^{-6}$C/m$^2$ are required to ensure that there is no appreciable effect on the trapping potential.}
    \label{tab:charge}
\end{table}

\subsection{Chiral Effects (or lack thereof)}
A consideration frequently required in atom traps using strong light confinement is that of chiral quantum optics \cite{lodahl_chiral_2017}. The emergent effect appears with waveguides of sub-$\lambda$ geometries ($w_\textrm{wg}<\lambda/n_\textrm{eff}$), where the local polarisation can be dependent to the propagation direction of the light, hence producing chiral fields. This can lead to inhomogeneous Zeeman broadening which arise from the vector light shifts which in turn make the trap state-dependent and spatially inhomogeneous. In fiber-based applications this is typically avoided by using counter-propagating modes and considering the light-shift Hamiltonian with respect to the polarisability tensor $\alpha ^{(2)}(\omega)$ \citep{goban_demonstration_2012,pache_magic-wavelength_2025}. 
In the system presented, modes are polarised along the TE0 direction, where $w_\textrm{wg}$ is significantly greater than $\lambda$; leading to minimal chiral effects to begin with and a trap that is proportionally much more localised (minimising spatial inhomogeneity). Thus, only trapping and transport modes are required within the PIC, and no modes are needed to remove the vector light shift; greatly simplifying the experimental complexity of the system.
\section{\label{sec:app}Applications}
\subsection{Clocks}
An important application for quantum atomic devices is that of compact clocks. In this context, the 778 nm two-photon $5S_{1/2}$ - $5D_{5/2}$ transition in $^{87}$Rb has been identified as promising candidate for distributed timing systems \citep{martin_compact_2018,di_gaetano_7781_2024}. In our case, because the trap lifetime exceeds the loading rate, the system could eventually provide a continuous source of cold atoms for clock interrogation. Our trapping scheme can be optimised for use in a clock by minimising AC Stark shifts from the transition and by taking advantage of the low Doppler shift incurred by interrogating a BEC. The AC stark shift causes the energy gap between transitions to change when optically trapped - the light induces a wavelength-dependent change in the atom's scalar polarisability for both the ground and excited state of the clock transition
\begin{equation}
\Delta\alpha(\omega)=\alpha_{5D_{5/2}}(\omega)-\alpha_{5S_{1/2}}(\omega).
    \label{eq:Clock_Delta_alpha}
\end{equation}
This change in polarisability manifests itself as a a frequency-shift to the transition, dependent on the trapping intensity and wavelengths of the modes

\begin{equation}
\Delta\nu=\sum_{n=r,b}\frac{\Delta \alpha(\omega)}{hc\epsilon_0}I_n(\textbf{r}_\textrm{trap}).
    \label{eq:clock_Delta_nu}
\end{equation}
Typically, atoms are trapped at so-called magic wavelengths where $\Delta\alpha(\omega_\textrm{magic})=0$, meaning no AC stark shift is present regardless of intensity. In our system magic wavelengths would need to be found for $^{87}$Rb that are both red- and blue-detuned from the trapping transition. Without this, it is still possible to balance the AC stark shifts of each trapping potential to produce a trap with low $\Delta\nu$ by rearranging Eq. \eqref{eq:clock_Delta_nu} to get the inequality
\begin{equation}
\frac{\Delta \alpha(\omega_b)}{\Delta \alpha(\omega_r)}=-\frac{I_r(\textbf{r}_\textrm{trap})}{I_b(\textbf{r}_\textrm{trap})}.
    \label{eq:clock_inequality}
\end{equation}
A trap with $\Delta T_\textrm{trap}=-0.5$ µK, $\Delta T_\textrm{barrier}=1.4$ µK  can be produced which provides a clock-shift of 5.4 Hz, which as a fraction of the two-photon 778 nm transition is $\frac{\Delta\nu }{2\nu_0}=7\times10^{-15}$. However an intensity stability $\Delta I/I\approx10^{-13}$ is required to maintain this scalar shift, meaning the system is highly sensitive to fluctuations. To minimise the instability, trapping wavelengths with smaller scalar polarisability can be chosen. For example, $\Delta \alpha(\lambda=738.56 \textrm{ nm})/\Delta\alpha(\lambda=720\textrm{ nm})\approx10^{-4}$ meaning the sensitivity is reduced by four orders of magnitude. Improvements in polarisability red-detuned from the trapping transition are harder to find, but improvements can equally be made by lowering $I_\lambda$ and exploring traps with different values of $\Delta T_\textrm{trap}$ and $\Delta T_\textrm{barrier}$, or releasing atoms from the trap during clock interrogation. 
Other considerations can be made for Doppler broadening - which is inherently very low for cold atoms, and minimised further in a BEC; in addition, the effect of the CP force will introduce a homogeneous shift to the transition's frequency. 

\subsection{Inertial Sensing}\label{sec:intertial sensing}

PIC-based systems like the ones demonstrated here can be utilised for compact, transportable atom interferometry for, e.g., inertial sensing. Here, we have designed our system so that it is compatible with schemes for phase-modulated, lattice-based, trapped-atom interferometry~\cite{weidner_experimental_2018,Holland_BBI}, but other efforts in modulated lattice interferometry are also amenable to the system presented here~\cite{weld,Gupta,Gauguet}. Additionally, if one removes the lattice and simply loads atoms into a one-dimensional waveguide structure, with limited confinement in the x-direction, it is possible to implement light-pulse atom interferometry protocols~\cite{rasel,hogan,sauer}, e.g., by sending relevant light pulses down the integrated photonic waveguide. It is worth noting, however, that one must consider the effect of wavefront aberrations when employing these methods~\cite{gaaloul}. 

In particular, when considering the shaken lattice interferometry protocol considered in Ref.~\cite{weidner_experimental_2018}, the protocol relies on atoms in relatively shallow lattices, as demonstrated in Figs.~\ref{fig:PIC_trap_parameters} and \ref{fig:PIC_saddle}. Typical phase modulation speeds are on the order of $\SI{100}{\kilo\hertz}$, which is possible with silicon nitride if other materials (like silicon or lithium niobate) are integrated within the platform~\cite{bowers}. As such, we predict that the protocols described in Ref.~\cite{weidner_experimental_2018} are readily implementable in our chip-scale system, although it is probable, at least initially, that sensitivities will suffer due to lower atom numbers in integrated traps relative to bulk optical lattices. 

Using a single, one-dimensional waveguide, the methods considered here will largely be limited to accelerometry, although similar proposals for rotation sensing have been recently developed~\citep{zhouSagnacTractorAtom2025}. However, given the compact size of the device, one can imagine an array of waveguides in orthogonal directions, which provides both options for relatively high-bandwidth atom interferometry (by running multiple PIC-based interferometers loaded from a single BEC) and multi-axis inertial sensing of accelerations and rotations. Given that chip-scale BECs have been produced at rates of \SI{2}{\hertz}~\cite{klempt}, such serial operation of a set of independent interferometers can, in principle, reach repetition rates in the tens to hundreds of interferometer cycles per second.

\hfill \break
\section{\label{sec:conc}Conclusion}

This work presents the operating principle of a PIC-based trap for a $^{87}$Rb BEC using a superposition of the evanescent fields of red- and blue-detuned fundamental and higher-order modes. We present a scheme for the entire experimental process, starting from cold atom generation in a gMOT to PIC trapping and manipulation. Trap parameters can be dynamically reconfigured, and the waveguide geometry adjusted to suit specific experimental requirements. The simulations presented demonstrate robustness to perturbations such as stray charges and laser power fluctuations, although these effects require experimental investigation for full quantification.

To progress this work, further detailed component- and PIC-level design is needed, along with exploration of potential applications to other atomic species and PIC platforms such as LiNbO$_3$. One must also develop and refine fabrication processes in order to enable a single, chip-scale device containing both a grating structure, photonic integrated circuits, and wires allowing for magnetic trapping and manipulation. Once demonstrated, and with appropriate wavelength choices minimising differential polarisabilities, the system could enable applications in inertial interferometry, compact clocks, and other domains within quantum science and engineering. 

During the preparation of this work, the authors became aware of work by Zhou et al.~\cite{zhouSagnacTractorAtom2025} that describes a similar method for atom trapping on a PIC. Their method focuses on inertial sensing using atoms trapped in ring waveguides with somewhat different loading conditions, and this work focuses on accelerometry and timekeeping, with the final goal of producing a self-contained ultracold atom sensor.

\begin{acknowledgments}
    The authors would like to thank Jon Pugh, Martin Cryan, Edward Deacon, Joe Cox, and Alex Clark for useful discussions. S.H. acknowledges funding from the EPSRC Centre for Doctoral Training in Quantum Engineering (EP/SO23607/1). C.W. acknowledges funding from the EPSRC UK Hub for Quantum Enabled Positioning, Navigation, and Timing  (EP/Z533178/1).
\end{acknowledgments}

\bibliography{bib_no_lang, references}

\end{document}